\documentclass[aps,pra,twocolumn,showpacs,groupedaddress]{revtex4}
\usepackage{graphicx}
\usepackage{amsmath}
\usepackage{longtable}
\usepackage{afterpage}

\begin{document}

\title{Energy and expectation values of the PsH system.  }
\author{J.Mitroy}
\affiliation{Faculty of Technology, Charles Darwin University, 
Darwin NT 0909, Australia}

\date{\today}

\begin{abstract}
Close to converged energies and expectation values for PsH are 
computed using a ground state wave function consisting of 1800 
explicitly correlated gaussians.   The best estimate of the PsH$^{\infty}$ 
energy was -0.789196740 hartree which is the lowest variational
energy to date.  The 2$\gamma$ annihilation rate for PsH$^{\infty}$ 
was $2.47178\times 10^{9}$ s$^{-1}$.  
 
\end{abstract}

\pacs{36.10.Dr, 36.10.-k, 31.15.Ar}

\maketitle

\vspace{1cm}

The calculation of positronium hydride (PsH) represents one of 
the simplest possibilities for studying mixed electronic
and positronic systems.  Since its stability was first identified
in 1951 by Ore \cite{ore51}, a variety of methods have
been applied to determine its structure.  These include 
variational calculations with Hylleraas type basis sets 
\cite{lebeda69,ho86,yan99a,yan99b}, 
variational calculations with explicitly correlated gaussians (ECGs) 
\cite{frolov97a,strasburger98,ryzhikh99a,usukura98},
quantum Monte Carlo methods 
\cite{yoshida96b,bressanini98a,bressanini98c,jiang98b,chiesa04a} 
and most recently the configuration interaction method 
\cite{bromley00a,mitroy01b,bromley02a,saito03a}.  The lowest variational 
energy for PsH$^{\infty}$ prior to the present article was that of Yan and Ho 
\cite{yan99a}.  Their largest calculation gave an energy of -0.7891967051 
hartree.  Bubin and Adamowicz used a 3200 dimension ECG basis to give 
an energy of -0.788870707 hartree for PsH$^{1}$ \cite{bubin04a}.   

\begingroup
\begin{table*}[bt]  
\caption[]{ \label{convergence} Behavior of some PsH
expectation values for a sequence of ECG type variational calculations 
of increasing size.  All quantities are given in atomic units with 
the exception of the $2\gamma$ annihilation rates which are in units 
of $10^9$ s$^{-1}$.  Some of the data for the earlier calculation 
\cite{ryzhikh99a} have not been published before, the data attributed 
to these calculations were computed using the same ECG basis.}
\vspace{0.2cm}
\begin{ruledtabular}
\begin{tabular}{lccccccccc}

$N$ & $\langle r_{\text{H}^{+}e^+} \rangle $ & $\langle r^2_{\text{H}^{+}e^-} \rangle $ & 
$\langle 1/r_{e^-e^-}\rangle $ &   
$\langle r_{e^+e^-}\rangle $ & $\langle \delta(e^--e^-) \rangle $ &  
$\langle \delta(\text{H}^{+}-e^+) \rangle $ &  $\Gamma$  & 
$\langle V\rangle/\langle T \rangle$ + 2 & $E$  \\ \hline  
750 \cite{ryzhikh99a} & 3.661596 & 7.812895 & 0.3705556 & 3.480249 
                      & $4.39845\times 10^{-3}$ & $1.63863\times10^{-3}$ & 2.46852 & $5.51\times10^{-7}$ & -0.789195993 \\   
900                   & 3.661613 & 7.812961 & 0.3705554 & 3.480263 
                      & $4.39321\times 10^{-3}$ & $1.63635\times10^{-3}$ & 2.46879 & $7.96\times10^{-7}$ & -0.789196542 \\   
1200                   & 3.661621 & 7.813024 & 0.3705550 & 3.480270 
                      & $4.38188\times 10^{-3}$ & $1.63153\times10^{-3}$ & 2.47129 & $2.21\times10^{-7}$ & -0.789196673  \\   
1500                   & 3.661624 & 7.813040 & 0.3705549 & 3.480271 
                      & $4.37628\times 10^{-3}$ & $1.62850\times10^{-3}$ & 2.47134 & $1.30\times10^{-7}$ & -0.789196718 \\   
1800                   & 3.661624 & 7.813046 & 0.3705549 & 3.480272 
                      & $4.37639\times 10^{-3}$ & $1.62828\times10^{-3}$ & 2.47178 & $7.3\times10^{-8}$ & -0.789196740 \\   
\multicolumn{4}{l}{Hylleraas $N = 5741$ \cite{yan99a,yan99b}} &  
                      &  &   & 2.47258 &          &  -0.789196705 \\     
\multicolumn{4}{l}{Hylleraas $N \to \infty$ extrapolation \cite{yan99a,yan99b}} &  
                      &  &   & 2.47264(2) &       &  -0.789196715(5) \\     
\end{tabular}
\end{ruledtabular}
\end{table*}
\endgroup

In this work, the stochastic variational method (SVM) is used to
construct a wave function with a lower energy that the best wave
function of Yan and Ho.  Indeed, the best SVM energy of -0.789196740 
hartree is even lower than the value estimated by Yan and Ho as the 
variational limit (e.g. -0.7891967147(42) hartree).  

The SVM used for this work has been described in a number of 
articles \cite{ryzhikh98e,varga97,suzuki98a} and only the 
briefest description is given here.  The SVM expands the wave 
function in a linear combination of ECGs.  Such basis functions 
have Hamiltonian matrix elements that can be computed very quickly 
and the energy is optimized by performing a trial and error search 
over the exponential parameters that define the basis. The SVM has 
been used to solve a number of many-body problems in different 
areas of physics \cite{ryzhikh98e,suzuki98a}. 

For the present set of calculations a basis containing 
1800 ECGs was used for the final calculation.  All the
optimizations of the ECG basis were done with the H mass 
set to $\infty$.  The annihilation rates given in Tables 
\ref{convergence} and \ref{properties} are proportional 
to the probability of finding an electron and a positron 
at the same position in a spin singlet state according to 
\begin{eqnarray}
\Gamma &=&  4\pi r_e^2 c \langle \Psi |
      \sum_{i} O^S_{ip} \delta( \mathbf{r}_i - \mathbf{r}_p ) | \Psi \rangle \\ 
       &=& 1.009394 \times 10^{11} \sum_{i} 
        \langle \delta( \mathbf{r}_i - \mathbf{r}_p ) \rangle_S \ ,
\end{eqnarray}
\cite{lee58a,ryzhikh99a,neamtan62}.  The sum is over the electron 
coordinates, the $\delta$-function expectation is evaluated in $a_0^3$, 
and $\Gamma$ is given numerically in s$^{-1}$.  The operator $O^S_{ip}$ 
is a spin projection operator to select spin singlet states for the $ip$ 
electron-positron pair.  

Table \ref{convergence} lists a number of expectation values obtained 
from a sequence of increasingly larger calculations.  The net energy 
improvement when the basis was increased from 900 to 1800 ECGS, 
while being subjected to additional optimization, was 1.98$\times 10^{-7}$ 
hartree.  It is worth noting that the energy of the largest calculation, 
namely -0.789196740 hartree, is lower than the previous best energy 
of Yan and Ho \cite{yan99a}, namely -0.7891967051 hartree. 
Yan and Ho examined the convergence pattern associated with their
sequence of increasingly larger calculations and estimated 
that the true energy was actually $9.6(4.2) \times 10^{-9}$ hartree
lower (e.g. -0.7891967147(42) hartree).  The present calculation 
indicates that the actual correction should have been more than
three times as large as that estimated by Yan and Ho.  Although the
sign of size of energy correction is not large, it is apparent
that the procedure used to determine the energy correction is 
faulty.  In Hylleraas calculations one typically does some sort 
of non-linear optimization to choose the exponential parameters
that give the minimum energy.  This has the unintended byproduct 
of distorting the convergence pattern of the energy and thus 
introducing large uncertainties in the extrapolation of the 
energy \cite{mitroy06c}.  This problem is probably more 
widespread than just the PsH calculation of Yan and Ho.  It
could occur whenever one extrapolates a sequence of energies 
while using a family of basis functions that are characterized by
a parameter which has been subjected to a non-linear optimization. 

\begin{table} [hb]  
\caption[]{ \label{properties} Properties of the PsH ground state.
Data are given for H assuming infinite mass.  All quantities are given 
in atomic units with the exception of the annihilation rates which are 
in units of $10^9$ s$^{-1}$. 
The positron and electron kinetic energy operators are written as
$T_+$ and $T_-$.  }
\vspace{0.2cm}
\begin{ruledtabular}
\begin{tabular}{lcc}

Property & Present SVM   &  SVM \cite{usukura98}  \\ \hline 
 $N$ & 1800   &    1600    \\ 
$\langle V\rangle/\langle T \rangle$ + 2 &
 $7.3 \times 10^{-8}$ &   $6\times10^{-7}$ \\   
$E$ & -0.789196740 & -0.789165554   \\ 
$\langle T_- \rangle$  &
  0.3261733  &    0.3261732       \\
$\langle T_+ \rangle$  &
  0.1368503 &    0.1368501      \\
$\langle r_{\text{H}^{+}e^-} \rangle $ & 
  2.311526  &     2.311525       \\
$\langle r_{\text{H}^{+}e^+} \rangle $ & 
  3.661624  &   3.661622       \\ 
$\langle r_{e^-e^-}\rangle $ & 
  3.574787  &   3.574783    \\
$\langle r_{e^+e^-}\rangle $ & 
  3.480272 &   3.480271     \\ 
$\langle 1/r_{\text{H}^{+}e^-} \rangle $ & 
  0.7297090 &   0.7297087   \\
$\langle 1/r_{\text{H}^{+}e^+} \rangle $ & 
  0.3474618   &  0.3474618    \\ 
$\langle 1/r_{e^-e^-}\rangle $ & 
  0.3705549   &  0.3705549     \\
$\langle 1/r_{e^+e^-}\rangle $ & 
  0.4184961  &    0.4184960 \\ 
$\langle r^2_{\text{H}^{+}e^-} \rangle $ & 
  7.813046  & 7.813015       \\
$\langle r^2_{\text{H}^{+}e^+} \rangle $ & 
  16.25453   &   16.25448     \\ 
$\langle r^2_{e^-e^-}\rangle $ & 
  15.87546 &    15.87538 \\ 
$\langle r^2_{e^+e^-}\rangle $ & 
  15.58427 &    15.58423  \\ 
$\langle 1/r^2_{\text{H}^{+}e^-} \rangle $ & 
  1.207067  & 1.207063        \\
$\langle 1/r^2_{\text{H}^{+}e^+} \rangle $ & 
  0.1721631  &   0.1721637     \\ 
$\langle 1/r^2_{e^-e^-} \rangle $ & 
  0.2139099 & 0.2139106       \\
$\langle 1/r^2_{e^{-}e^+} \rangle $ & 
  0.3491440  &   0.3491428     \\ 
$\langle \delta(\text{H}^{+}-e^-) \rangle $ & 
  0.177279 &    0.177186    \\ 
$\langle \delta(\text{H}^{+}-e^+) \rangle $ & 
   $1.62828 \times 10^{-3}$ &   $1.63857 \times 10^{-3}$   \\ 
$\langle \delta(e^--e^-) \rangle $ & 
  4.37639$\times 10^{-2}$  & 4.3867$\times 10^{-3}$   \\ 
$\langle \delta(e^+-e^-) \rangle $ & 
  0.0244877  &   0.024461   \\ 
$\Gamma$ & 
  2.47178 &  2.46909  \\
\end{tabular}
\end{ruledtabular}
\end{table}

The coalescence matrix elements, $\langle \delta(e^--e^-) \rangle$   
and $\langle \delta(\text{H}^{+}-e^+) \rangle $ were more 
sensitive to the increase in basis size than any other 
quantity.  This sensitivity is due to the fact that the wave 
function amplitude between two repelling particles is expected
to be small at their coalescence point and the ECG functional 
form  is not the natural choice to describe the behavior 
of the relative wave function for two strongly repelling 
particles.   With respect to the more physically interesting 
observables, the annihilation rate, $\Gamma$ varied most as 
the basis dimension was increased.  But, the increase in 
$\Gamma$ was just larger than 0.1$\%$ when the basis was increased 
from 900 to 1800.        

A comprehensive set of the best set of expectation values are listed 
in Table \ref{properties}.  They are compared with the results of another, 
but completely independent, large basis SVM calculation \cite{usukura98}.  
The expectation value for the virial theorem 
$\langle V \rangle /\langle T \rangle$ 
provides an estimate of the wave function accuracy and the 
deviation of $\langle V \rangle /\langle T \rangle$ from -2 
was only $7.3 \times 10^{-8}$ hartree.    

The energies of the different mass variants of PsH 
were computed by rediagonalizing the Hamiltonian with the 
same basis but with $m_{\text{H}^1}$ set to 1836.1527 $m_e$,  
$m_{\text{H}^2}$ set to 3670.483 $m_e$ and 
$m_{\text{H}^3}$ set to 5496.899 $m_e$.   The energies of 
PsH$^{1}$, PsH$^{2}$ and PsH$^{3}$ were -0.788870618 hartree, 
-0.789033556 hartree and -0.789087767 hartree respectively.   
The energy of the 3200 ECG wave function of Bubin and Adamowicz \cite{bubin04a}  
for PsH$^{1}$ was -0.788870707 hartree, which is $1.0 \times 10^{-7}$ 
hartree below the present energy.   

To summarize, a close to converged binding energy is reported
for the PsH$^\infty$ ground state.  The present energy is 
$2.5 \times 10^{-8}$ hartree lower than the estimated variational 
limit of Yan and Ho.   The procedure by Yan and Ho to estimate 
the variational limit probably tends to underestimate the size 
of the necessary energy correction.   

Although the present energy is better than that of Yan and Ho, 
this does not necessarily mean that the present SVM annihilation 
rate is more accurate.  Any basis of ECGs  
(which cannot satisfy the exact inter-particle cusp conditions) 
will have a tendency to underestimate the electron-positron
coalescence matrix element.   Table \ref{convergence} shows a 
consistent increase in $\Gamma$ as the size of the calculation 
in increased.  
 
This work was supported by a research grant from the Australian
Research Council.  The authors would like to thank S Caple 
for access to additional computer facilities. 

%\bibliography{positron}

\end{document}